\documentclass[prd,aps,eqsecnum,floatfix,nofootinbib,preprint,tightenlines]{revtex4}

\usepackage{latexsym}
\usepackage{graphicx}
\let\inodot=\i

\def\d{\delta}
\def\i{\iota}

\def\p{\pi}                     % Also, \varpi
\def\cbo{{\,\raise-.15ex\Sc [\,}}                       % curly "
\def\gtap{\raisebox{-.4ex}{\rlap{$\sim$}} \raisebox{.4ex}{$>$}}   % > or ~
\def\ddt#1{{\buildrel {\hbox{\LARGE .\kern-2pt.}} \over {#1}}}% double dot-over
\def\ie{\mbox{\it i.e.}}

\def\tr{{\rm tr}\,}

\long\def\symbolfootnote[#1]#2{\begingroup%
\def\thefootnote{\fnsymbol{footnote}}\footnote[#1]{#2}\endgroup}

\long \def \blockcomment #1\endcomment{}

\def\ie{{\it i.e.}}
\def\ceef{{\it cf.}}

\def\tr{{\rm tr}}

\def\vp{{\vec p}}
\def\vq{{\vec q}}
\def\vk{{\vec k}}
\def\vn{{\vec n}}
\newcommand{\pref}[1]{(\ref{#1})}

\def\ansatz{{\it ansatz}}

\begin{document}
\hyphenation{fer-mio-nic per-tur-ba-tive pa-ra-me-tri-za-tion
pa-ra-me-tri-zed a-nom-al-ous}

\renewcommand{\thefootnote}{$*$}

\hfill BI-TP 2012/37

\hfill HU-EP-12/26

\hfill SFB/CPP-12-63

\begin{center}
\vspace*{10mm}
{\large\bf Excited-state contribution to the
axial-vector and pseudo-scalar correlators with two extra pions}
\\[12mm]
Oliver B\"ar$^{a,b}$ and Maarten Golterman$^c$
\symbolfootnote[2]{Permanent address: Department of Physics and Astronomy
San Francisco State University, San Francisco, CA 94132, USA.}
\\[8mm]
{\small\it
$^a$Institut f\"ur Physik,
\\Humboldt Universit\"at zu Berlin,
12489 Berlin, Germany}
\\[5mm]
{\small\it
$^b$Institut f\"ur Physik,
\\Universit\"at Bielefeld,
33615 Bielefeld, Germany}
\\[5mm]
{\small\it
$^c$Institut de F\'\inodot sica d'Altes Energies, Universitat Aut\`onoma de Barcelona,\\ E-08193 Bellaterra, Barcelona, Spain}
\\[10mm]

Abstract\\[5mm]

\begin{minipage}{15cm}
We study multi-particle state contributions to the QCD two-point functions of the axial-vector and pseudo-scalar quark bilinears in a finite spatial volume. For sufficiently small quark masses  one expects three-meson states with two additional pions at rest to have the lowest total energy after the ground state. 
We calculate this three-meson state contribution using chiral perturbation theory. We find it to be strongly suppressed and too small to be seen 
in present-day lattice simulations. 

\end{minipage}
\\[2mm]
\end{center}

\renewcommand{\thefootnote}{\arabic{footnote}} \setcounter{footnote}{0}

\newpage
\section{\label{Intro} Introduction}

Hadron spectroscopy is one of the prime applications of lattice QCD simulations. 
Recent results for the light hadron spectrum in 2+1 flavor simulations show an  agreement with the experimentally measured values to an accuracy of a few percent \cite{Ukita:2007cu,Durr:2008zz}. 
The computation of the excited-state spectrum is much more complicated, and the numerical results are not as satisfactory as those for stable hadrons 
\cite{Dudek:2010wm,Bulava:2010yg}. Still,  steady progress in excited-state spectroscopy has been made over the last years (for a recent review see Ref.~\cite{Bulava:2011uk}).

One feature of unquenched lattice simulations is the presence of multi-particle states in the correlation functions measured to obtain the spectrum. In particular, with the up and down quark masses getting closer to their physical values, one may expect three-particle states with two additional pions at rest to contribute dominantly after the ground state in a given channel. Evidence for this expectation has been reported, for example,  in Refs.~\cite{DelDebbio:2006cn,DelDebbio:2007pz}. 
There results for the two-point correlation function $C(t)$ of the pseudo-scalar density were analyzed, as a function of the time separation $t$. The data were well described by an \ansatz\ involving two exponentials,
\begin{equation}
\label{PPansatz}
C(t) = c_0\,e^{-Mt} +c_1\,e^{-M^{\prime}t}\ ,\nonumber
\end{equation}
with $M^{\prime}=M+2m_{\pi}$ and the other three parameters determined by the fit.\footnote{To be precise, what was analyzed is the effective mass,
equal to minus the time derivative of $\log C(t)$.}  
A significant dependence of the correlator on the sea-quark mass was observed, supporting the interpretation of the second term in Eq.~(\ref{PPansatz}) as a three-particle contribution. 

The question here is whether one really sees a three-particle state and not a genuine one-particle excited state.
In fact, the results in Refs.~\cite{DelDebbio:2006cn,DelDebbio:2007pz} are somewhat surprising. On theoretical grounds one expects the coefficient $c_1$ of a three-particle state to be suppressed by two powers of the (spatial) lattice volume and therefore to be rather small.
In principle, this volume dependence of the three-particle state can be monitored, providing a strong check for the three-particle state hypothesis. However, this would require lattice data at various volumes, 
and these are often not available. 

The main observation in this paper is that the correlation function of the pseudo-scalar density and the axial-vector current can be calculated reliably in chiral perturbation theory (ChPT) \cite{Weinberg:1978kz,Gasser:1983yg,Gasser:1984gg}. 
ChPT is expected to give good estimates for the coefficients $c_0$ and $c_1$ for small pion masses where the exponential suppression of the three-particle state contribution in the correlator is less strong. Moreover, to leading order (LO) in the chiral expansion, the coefficients $c_0$ and $c_1$ depend only on the pseudo-scalar masses and decay constants, hence they are related. It turns out that, in case of both the axial-vector and pseudo-scalar correlation functions,   
the coefficient $c_1$ of the three-particle state is completely determined in terms of $c_0$ and the Goldstone-meson masses, reducing the number of fit parameters in Eq.~(\ref{PPansatz}) from three to two. 

Our results show that the ratio $c_1/c_0$ is very small, even smaller than a
naive analysis would suggest.
 For typical pion masses and lattice volumes of present lattice simulations it is of order $10^{-4} - 10^{-3}$. Therefore, the size of the three-pion contribution is smaller than typical statistical errors in the lattice data and it plays no role in practice. This seemingly negative result can actually be viewed as a positive result, if one  is interested in ``genuine'' excited QCD states, and not in the multi-particle state contaminations.

This paper is organized as follows.
After a brief more general discussion of correlators in a finite volume
(Sec.~\ref{secQCD}), we present our results in Sec.~\ref{ChPT}, with
subsections~\ref{BasicDef} and\ \ref{2piCont} containing technicalities.
Section~\ref{sec:data} confronts our results with lattice data, and Sec.~\ref{sec:concl}
contains our conclusions and some further discussion.

\section{\label{secQCD} QCD correlators in a finite spatial volume}
Throughout this article we will consider QCD in a finite spatial box with length $L$ in each direction and periodic boundary conditions. The euclidean time extent, however, is assumed to be infinite. This choice  simplifies our calculations, and is justified as a good approximation for the many lattice QCD simulations with $T\gtap 2L$.

We are interested in the two-point correlation functions of the (zero component of the) axial-vector current and the pseudo-scalar density:
\begin{eqnarray}
C^a_{AA}(t)& =& \int_{L^3} {\rm d}^3{{x}}\, \langle  A_0^a(\vec{x},t) A_0^a(0,0)\rangle\ ,\label{ACorr}\\
C^a_{PP}(t) &= &\int _{L^3}{\rm d}^3{{x}}\, \langle  P^a(\vec{x},t) P^a(0,0)\rangle\ ,\label{PCorr}
\end{eqnarray}
where the flavor non-singlet current and density are defined as usual by
\begin{eqnarray}
\label{OpQCD}
A_{\mu}^a &=&  \overline{\psi} \gamma_{\mu}\gamma_5T^a\psi\ ,\qquad
P^a \,=\,\overline{\psi} \gamma_5 T^a\psi\ .
\end{eqnarray}
Unless stated otherwise, we consider 2+1 flavor QCD with degenerate up and down quarks with mass $m$ and a heavier strange quark with mass $m_s$.
Therefore, the flavor index $a$ runs from $1$ to $8$, and the $SU(3)$ group generators $T^a$ are chosen equal to $\lambda^a/2$, with $\lambda^a$ the usual Gell-Mann matrices. 

With our choice for the quark masses the correlators obey isospin symmetry, so it is sufficient to study the two cases $a=1$ and $a=4$.   
In addition, the correlators \pref{ACorr} and \pref{PCorr} are not independent because the axial-vector current and the pseudo-scalar density are related by the partially conserved axial-vector current (PCAC) relations, 
\begin{equation}
\label{pcac1}
\partial_{\mu}A_{\mu}^1 = 2mP^1\ ,\qquad\partial_{\mu}A_{\mu}^4 = (m+m_s)P^4\ .
\end{equation}
Using these relations the pseudo-scalar correlator can be obtained from the axial-vector one by taking two time derivatives.  In the following we will thus focus on $C^a_{AA}(t)$, although much of the following discussion applies to $C^a_{PP}(t)$ as well.

The integration over the spatial volume in Eqs.~(\ref{ACorr}) and~(\ref{PCorr}) projects on states with zero total momentum. Hence, the dominant contribution comes from the single-particle state with the particle being the appropriate pseudo-scalar meson at rest.  For the axial-vector correlator we obtain
\begin{equation}
\label{spcontr}
C^a_{AA,1\pi}(t)=
\frac{1}{2m_a}\;e^{-m_a|t|}\;|\langle 0|A^a_0(0)|\pi_a(\vp=0)\rangle|^2\ \,.
\end{equation}
In order to keep the notation simple we will often refer to all pseudo-scalars as pions and denote them by $\pi_a$, even though for $a>3$ they correspond to the kaons and the eta. The same applies to the corresponding masses $m_a$, which will denote either the pion, kaon or eta mass, depending on the value of the index $a$. In addition, we will always assume that $t>0$, so that $|t|=t$.

Note that for Eq.~(\ref{spcontr}) we assumed the standard normalization of one-particle states,
\begin{equation}
\label{norm}
\langle\pi_a(\vq)|\pi_b(\vp)\rangle=\delta_{ab} L^3\;2E_{\vp,a}\;\d_{\vp,\vq}\ ,
\end{equation}
in a finite volume.
Here $E_{\vp,a}=\sqrt{\vp^{\,2}+m_a^2}$ is the energy of the pion $\pi_a$, and
$\d_{\vp,\vq}$ is a Kronecker delta. With this convention the state $|\pi_a(\vp)\rangle$ and all other one-particle states have mass dimension $-1$. Consequently, the matrix element 
\begin{equation}
\label{Deffa}
\langle 0|A^a_0(0)|\pi_a(\vp=0)\rangle\equiv m_a f_a\ ,
\end{equation}
defining the decay constant $f_a$ in the usual way, and the correlator $C^a_{AA,1\pi}$ have their standard mass dimensions of 2 and 3, respectively. 

The axial-vector current excites other states with the same quantum numbers as well. The contribution of an excited pion $\pi_a^{\prime}$ has the same form as Eq.~(\ref{spcontr}) with the appropriate mass $m'_a\gg m_a$. However, for sufficiently small pion masses one expects the three-particle state with two additional pions to have a smaller energy than this excited state. For this three-particle contribution one finds
\begin{eqnarray}
\label{tpcontr}
\hspace{-0.7cm}C^a_{AA,3\pi}(t)&\!\!\!=\!\!\!&\frac{1}{L^6}\;\sum_{\vp,\vq,\vk}\d_{\vp+\vq+\vk,0}\;\frac{1}{8E_{\vp,a} E_{\vq,b} E_{\vk,b}}\,
|\langle 0|A_0(0)|\pi_a(\vp) \pi_b(\vq) \pi_b(\vk) \rangle|^2\,
e^{-E_{\rm tot}t}\ ,
\end{eqnarray}
where $E_{\rm tot}$ is the total energy of the state, and we did not show a possible wave-function renormalization. For weakly interacting pions $E_{\rm tot}$ equals approximately the sum $E_{\vp,a}+E_{\vq,b}+E_{\vk,b}$ of the individual pion energies and we expect the wave-function renormalization to be close to one. 

In a finite volume with periodic boundary conditions the momenta $\vp=2\p\vn/L$ are
quantized, with $\vn$ having integer-valued components. If $2\p/(m_\p L)$ is not much smaller than one, states with at least one of the pions having a non-zero
momentum will be signficantly more energetic than those in which all
three particles are at rest, and it is sufficient to keep the latter
contribution only, \ie,
\begin{equation}
\label{tpzerop}
C^a_{AA,3\pi}(t)=\frac{1}{L^6}\;\frac{1}{8m_a m_b^2}\;
|\langle 0|A_0(0)|\pi_a(\vp=0) \pi_b(\vq=0) \pi_b(\vk=0) \rangle|^2\;
e^{-(m_a + 2m_b)t}  \ .
\end{equation}
Here the index $b$ is restricted to the values $1$, $2$, and $3$, such that the three-particle state is indeed a ``pseudo-scalar-plus-two-pion-state.'' 
This state is also lighter than a three-particle state with two additional kaons. Note that the dimensions in Eq.~(\ref{tpzerop}) come out right: The matrix element on the right hand side is dimensionless, so the left hand side has dimension 3, as the leading one-particle contribution.

The three-particle contribution is suppressed by $1/L^6$. This factor is usually taken to argue that the three-particle contribution in the correlator is small despite the fact that it might be the most important excited-state contribution according to the exponential suppression. To make this a little more quantitative consider the ratio of the three-particle and the one-particle contributions, assuming on dimensional grounds that the matrix element in Eq.~(\ref{tpzerop}) is of the order $m_\pi/f_\pi$ (for $a=1$). We thus estimate
\begin{equation}
\label{guess}
\frac{C^1_{AA,3\pi}(t)}{C^1_{AA,1\pi}(t)} \approx \frac{1}{4(f_\pi L)^4(m_{\pi}L)^2} \;e^{-2m_{\pi}t}\ .
\end{equation}
This is a very small number indeed. Typical values for 
$m_{\pi}L$ in present day lattice simulations are of order 4, while
$f_\pi L$ is of order 1.5, leading to a prefactor of the order $10^{-2}$ or
$10^{-3}$. 
In the next section more precise results will be obtained within ChPT.

\section{\label{ChPT} The correlators in ChPT}
We now turn to a calculation of the ratio~(\ref{guess})
for various mesonic ground states in three-flavor ChPT.   We will first set up the
framework in Secs.~\ref{BasicDef} and \ref{2piCont}, after which we will present results
for the axial-vector correlator in Sec.~\ref{AAresults} and for the pseudo-scalar
density correlator in Sec.~\ref{PPresults}.   In Sec.~\ref{DsMeson} we extend
our results to include the case of a ground-state meson made out of two
different quarks both with masses equal to the strange quark mass, which
is helpful for interpreting some of the results of Refs.~\cite{DelDebbio:2006cn,DelDebbio:2007pz}.   

\subsection{\label{BasicDef} Basic definitions}
We consider standard finite-volume ChPT for 2+1 flavors in euclidean space-time \cite{Gasser:1984gg,Gasser:1987zq}. The leading-order (LO) chiral lagrangian we use reads
\begin{equation}
\label{Lag}
{\cal L}_{\chi} = \frac{1}{4}f^2\; \tr\left({\partial_{\mu}\Sigma \partial_{\mu}\Sigma^{\dagger}}\right) - \frac{1}{2}f^2B\; \tr\left({M^{\dagger}\Sigma+\Sigma^{\dagger}M}\right)\ ,
\end{equation}
where the field $\Sigma$ is the standard $SU(3)$-valued chiral field containing the pseudo-scalar fields
\begin{equation}
\label{Sigma}
\Sigma = \exp\left(\frac{2i}{f}\sum_{a=1}^8 \pi_aT^a\right)\,.
\end{equation}
 $B$ and $f$ are the familiar low-energy coefficients (LECs) related to the chiral condensate and the decay constant in the chiral limit. To LO we have $f=f_{\pi}=f_K$, and our conventions correspond to $f_{\pi}= 92.2$~MeV.
The mass matrix $M$ in the mass term of the chiral lagrangian is, according to our setup in the previous section, given by
\begin{equation}
\label{Mmatrix}
M={\rm diag}(m,m,m_s)\ .
\end{equation}
Expanding the chiral lagrangian in powers of the pion fields we obtain the well-known LO pseudo-scalar masses
\begin{equation}
m_{a}^2=
\left\{\begin{array}{rclcl}
2Bm & \equiv&  m_{\pi}^2\ , & \quad & a=1,2,3\ ,\\
B(m+m_s) & \equiv & m_K^2\ , & \quad & a=4,\ldots,7\ ,\\
\frac{2}{3}B(m+2m_s) & \equiv& m_{\eta}^2\ ,& \quad & a=8\ .
\end{array}
\right.\label{LOMasses}
\end{equation}
With the spatial momentum discretized in a finite volume,
the (euclidean) pseudo-scalar propagators are given by
\begin{equation}\label{PropMomSpace}
G_{ab}(\vec{x}-\vec{y},t - t') = \langle \pi_a(\vec{x},t)\pi_b(\vec{y},t')\rangle=\delta_{ab} \frac{1}{L^3}\sum_{\vec{p}} e^{i\vec{p}\cdot(\vec{x}-\vec{y})} 
\frac{1}{2E_{\vec{p},a}} \,e^{-E_{\vec{p},a}|t-t'|}\ ,
\end{equation} 
with the energy $E_{\vec{p},a}=\sqrt{\vec{p}^2 + m_a}$.

The axial-vector current and the pseudo-scalar density corresponding to the definitions in Eq.~(\ref{OpQCD}) are
\begin{eqnarray}
A_{\mu}^a& =& \frac{1}{2}f^2\;{\rm tr}\Big(T^a(\Sigma^{\dagger}\partial_{\mu}\Sigma - \Sigma\partial_{\mu}\Sigma^{\dagger})\Big) \ ,
\label{Avect}\\
P^a &=& \frac{1}{2}f^2B\; {\rm tr} \Big(T^a(\Sigma - \Sigma^{\dagger})\Big)\ ,
\label{Pscalar}
\end{eqnarray}
Since we will be working to leading order (LO) in this article, we will only need
these LO expressions.   
Higher-order terms, involving Gasser-Leutwyler (GL) coefficients \cite{Gasser:1984gg}, will thus not be needed. 

Expanding Eqs.~(\ref{Avect}) and~(\ref{Pscalar}) to first order in the pion fields and substituting into Eqs.~(\ref{ACorr}) and~(\ref{PCorr}) it is simple to obtain 
\begin{eqnarray}
C^a_{AA}(t)& =&\phantom{-} \frac{1}{2}f^2m_a \,e^{-m_a t}\ ,\label{ACorrLO} \\
C^a_{PP}(t) &= &- \frac{f^2B^2}{2m_a}\, e^{-m_a t}\label{PCorrLO}\ .
\end{eqnarray} 
As already mentioned, these correlators are not independent, but related by the PCAC relations~(\ref{pcac1}), which can be used to show that
\begin{equation}
\label{relcorr}
C^a_{PP}(t)=-\frac{B^2}{m_{a}^4} \;\partial^2_t C^a_{AA}(t)\ ,
\end{equation}
which is indeed satisfied by Eqs.~(\ref{ACorrLO}) and~(\ref{PCorrLO}).
The PCAC relations holds to all orders in the chiral expansion and for all $n$-particle contributions.  We will use it in section \ref{PPresults} to calculate the three-pion contribution in $C^a_{PP}(t)$. 

A graphical representation of these LO correlation functions is given in Fig.\ \ref{fig:diagrams}a. The two open squares represent the one-pion terms for either the current or the density, which (after contraction of the fields) result in the propagator represented by the line.

\subsection{\label{2piCont} Multi-meson state contribution to the correlators}
\begin{figure}[tp]
\begin{center}
\includegraphics[scale=0.45]{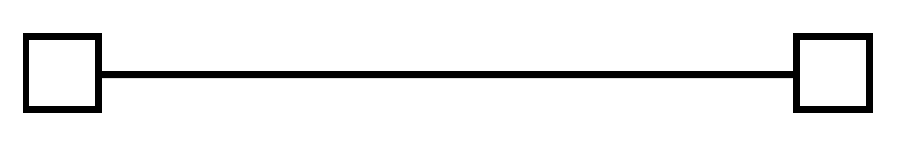}\\
a)\\[3ex]
\includegraphics[scale=0.45]{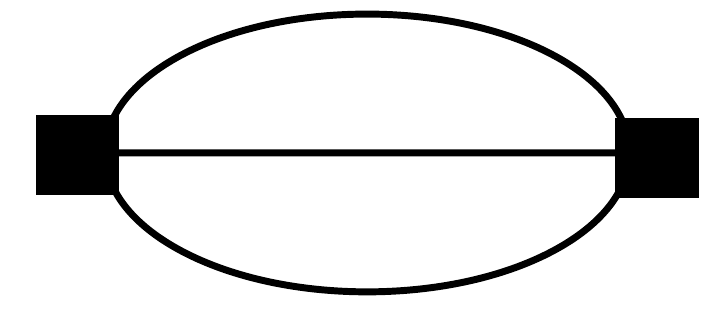}\\
b)\\[3ex]
\includegraphics[scale=0.45]{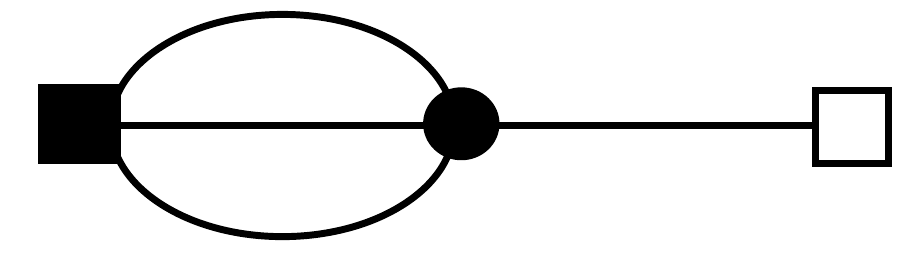}\hspace{1cm}\includegraphics[scale=0.45]{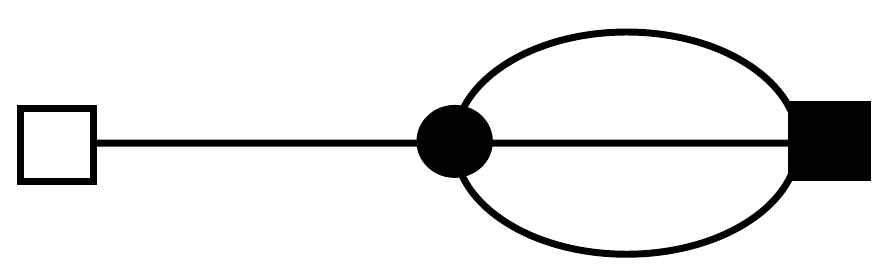}\\
c)\hspace{5cm} d)\\[3ex]
\includegraphics[scale=0.45]{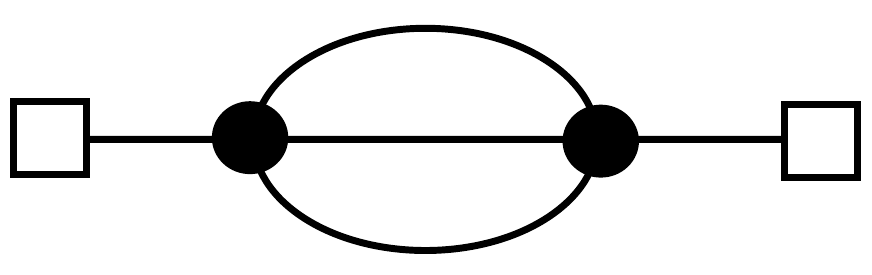}\\
e)
\caption{Feynman diagrams for the axial-vector correlation function. The squares represent the axial-vector current at times $t$ and $0$, where the open and solid squares denote the one-pion and three-pion terms in the current, respectively. The circles represent a vertex insertion at an intermediate space time point; an integration over these points is implicitly assumed. The lines represent propagators in position space. At least two of the three lines connecting a solid square or circle are pion propagators in our calculation.}
\label{fig:diagrams}
\end{center}
\end{figure}

The results~(\ref{ACorrLO}) and~(\ref{PCorrLO}) fall off exponentially with the pion and kaon mass for $a=1$ and $a=4$, respectively. The diagrams in Figs.~\ref{fig:diagrams}b to \ref{fig:diagrams}e lead to contributions with the anticipated exponential decay with rates $3m_{\pi}$ for $a=1$  and $m_K+2m_{\pi}$ for $a=4$. In these diagrams the solid square represents the three-pion terms of the current or density. These terms contain three pseudo-scalar fields leading to three propagator lines emanating from them. 
If two of the three propagators correspond to pion propagators the diagram leads to an exponential with exponent $-(m_a+2m_{\pi})t$.
Similarly, the circle represents the insertion of an interaction vertex stemming from terms with four pseudo-scalar fields in the chiral lagrangian. The vertex includes an integration over the intermediate space-time point where the vertex is placed. Also diagram~\ref{fig:diagrams}e with two LO currents or densities placed at times $t$ and $0$ and two inner pion propagators has a contribution with time dependence $\exp[-(m_a+2m_{\pi})t]$.

Note that all diagrams in Figs.~\ref{fig:diagrams}b to \ref{fig:diagrams}e are of the same order in the chiral expansion. In order to see this, recall the power counting of finite-volume ChPT in the $p$-regime, for which
\begin{equation}
\label{pregcount}
m_{\rm quark}\;\sim O(p^2)\ , \;\;\;\;1/L\sim\partial_\mu\; \sim O(p)\ ,\;\;\;\; \pi_a \;\sim O(p)\ .
\end{equation}
With these counting rules diagram~\ref{fig:diagrams}a counts as order $p^4$, since each axial-vector current involves one pion field and one derivative. Diagram~\ref{fig:diagrams}b counts as order $p^8$, since here each current consists of three pseudo-scalar fields and one derivative. For diagrams~\ref{fig:diagrams}c to \ref{fig:diagrams}e with one or two vertex insertions, note that each vertex involves four pseudo-scalar fields associated with either two derivatives or one power of $m_{\rm quark}$. This is of order $p^6$. Taking into account the integration over space time, which counts as order $p^{-4}$, a vertex insertion effectively counts as order $p^{2}$. Hence, the diagrams in Figs.~\ref{fig:diagrams}c--e count as order $p^8$ as well. Similar arguments can be made for the pseudo-scalar correlator, and one finds again that the diagrams in Figs.~\ref{fig:diagrams}b--e are all of the same order in the chiral expansion. Diagrams involving GL coefficients $L_i$, stemming from higher-order terms in the current and the chiral lagrangian \cite{Gasser:1984gg}, are at least of order $p^2$ higher since they contain either two more partial derivatives or one extra power of the quark mass.  

We are interested in the contributions to the correlator that decay with $3m_{\pi}$ ($a=1$) or $m_K+2m_{\pi}$ ($a=4$). This allows us to simplify the calculation by ignoring many terms in the expressions for the axial-vector current and the interaction vertices that do not contribute. For the correlator $C^a_{AA,3\pi}$ we need to keep only those terms in $A_0^a$ that involve one $\pi_a$ field and two light fields, \ie, fields with flavor index  $1,2$ or $3$. For the interaction vertices we need to keep the vertices with at least on $\pi_{a}$ in order to contract with a field from the current. Two more fields need to be light fields in order to end up with two pion propagators in the intermediate lines in the diagrams. 
With these restrictions in mind we can work with the following (incomplete!) currents:
\begin{eqnarray}
A_{\mu,{\rm NLO}}^1& =&\frac{2i}{3f} \Big( \pi_1 \pi_2 \partial_{\mu}\pi_2  - \partial_{\mu}\pi_1\pi_2^2  \,+\, (2\rightarrow 3)\Big)+\dots\ ,\label{A1NLO}
\\
A_{\mu,{\rm NLO}}^4& =& \frac{i}{6f} \Big( \pi_4\pi_b \partial_{\mu}\pi_b - \partial_{\mu}\pi_4 \pi_b^2\Big)+\dots\ .\label{A4NLO}
\end{eqnarray}
where the dots indicate the terms not needed for our calculation.
In the first line one has to add the same terms with the replacement $2\rightarrow 3$ for the flavor index, as indicated.
In the second line a sum over the light flavors is implied ($b=1,2,3$).

The relevant interaction vertices needed for diagrams~\ref{fig:diagrams}b--e read
\begin{eqnarray}
 v^1& =& -\frac{1}{24f^2} \Bigg( -4\pi_1^2(\partial_{\mu}\pi_2)^2 +8\pi_1\partial_{\mu}\pi_1\pi_2\partial_{\mu}\pi_2 - 4(\partial_{\mu}\pi_1)^2 \pi_2^2 \,+\, (2\rightarrow 3)\nonumber \\
 && \hspace{1.6cm}+\, m_{\pi}^2 (\pi_1^4 + 2\pi_1^2\pi_2^2 \,+ 2\pi_1^2\pi_3^2)\Bigg)+\dots\ ,\label{PionV}\\
v^4&=& -\frac{1}{24f^2}\Bigg(\pi_4^2 (\partial_{\mu}\pi_b)^2 - 2 \pi_4\partial_{\mu}\pi_4(\pi_b\partial_{\mu}\pi_b) + (\partial_{\mu}\pi_4)^2\pi_b^2\nonumber \\
&& \hspace{1.6cm} +\,  (m_K^2 + m_{\pi}^2)\pi_4^2\pi_b^2\Bigg)+\dots\ ,\label{KaonV}
\end{eqnarray}
where again the dots stand for terms not needed for our calculation,
and in $v^4$ a sum over $b=1,2,3$ is implied.  

\subsection{\label{AAresults} Results for the axial-vector correlator}
The diagrams in Fig.~\ref{fig:diagrams} are straightforwardly calculated. Two remarks, however, seem to be appropriate. 
First, recall that we are interested only in the contribution with all particles at rest. Hence, although the diagrams \ref{fig:diagrams}b--e  are two-loop diagrams, we do not need to perform the sum over internal momenta, \ceef\ Eq.~(\ref{tpcontr}).  Obviously, this simplifies the calculation significantly. 
Second, note that diagram \ref{fig:diagrams}e also contains contributions with a time dependence proportional to $t\exp(-m_a t)$. These contributions result in the standard renormalization of the pseudo-scalar masses, and can simply be omitted for our purposes.

For the pion correlator ($a=1$) we find
\begin{equation}
\label{C1AAres}
C^1_{AA}(t) =  \frac{1}{2}f_{\pi}^2m_{\pi}\,e^{-m_{\pi}t}\left[1+ \frac{5}{512 (f_{\pi}L)^4(m_{\pi} L)^2}\,e^{-2m_{\pi}t}\right] \ .
\end{equation}
In writing Eq.~(\ref{C1AAres}), we have made the replacement $f\rightarrow f_{\pi}$ in the three-pion contribution, since the difference is of higher order in the chiral counting. Note that the prefactor of the three-pion contribution is of the same form as in Eq.~(\ref{guess}).  There is a numerical factor $5/512$ in the ChPT result which leads to an even larger suppression of the three-pion
contribution in comparison with our naive estimate~(\ref{guess}).

For the kaon correlator ($a=4$) we obtain 
\begin{equation}
\label{C4AAres}
C^4_{AA}(t) = 
\frac{1}{2}f_K^2m_{K}\,e^{-m_{K}t}\Bigg[1+ \frac{1}{24 (f_K L)^4(m_{\pi}L)^2} \,h_1\left(\frac{m_{\pi}}{m_{K}}\right) e^{-2m_{\pi}t}\Bigg]\ ,
\end{equation}
where the function $h_1(x)$ is given by
\begin{equation}
\label{Defh}
h_1(x)= \left(1-x + \frac{1+2x^2}{4(1+x)}\right)^2\ .
\end{equation}
As in Eq.~(\ref{C1AAres}), we have here substituted $f\rightarrow f_{K}$. However, to the order we are working we could equally well substitute $f\rightarrow f_{\pi}$ or  replace $(f_KL)^4$ by $(f_KL)^2(f_{\pi}L)^2$ in the denominator of Eq.~(\ref{C4AAres}).   The difference is of higher order in the
chiral expansion than we consider here.

\begin{figure}[tp]
\begin{center}
\includegraphics[scale=0.7]{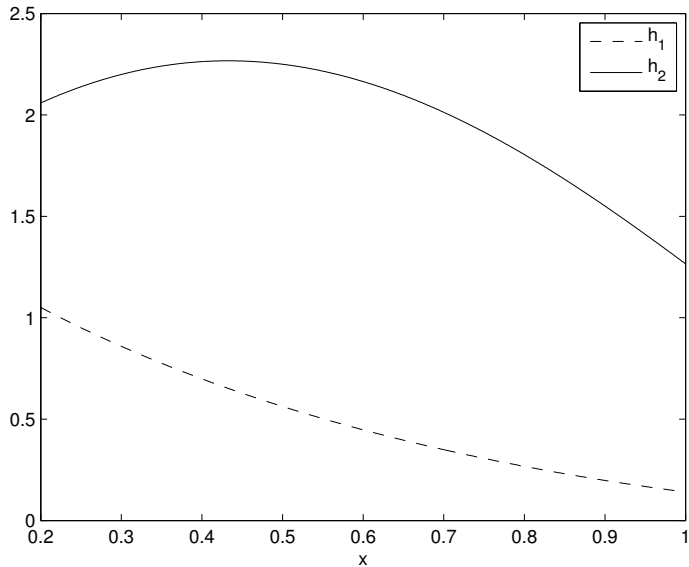}
\caption{The functions $h_i(x), \,i=1,2$ that enter the results for $C^4_{AA}(t)$ and $C^4_{PP}(t)$.}
\label{fig:hfunc}
\end{center}
\end{figure}

The function $h_1(x)$ is a well-behaved function of the mass ratio $m_{\pi}/m_K$, monotonically falling from about 0.9 at the physical mass ratio $x = 139/493\approx 0.3$ to about $1/7$ at equal masses (see Fig.\ \ref{fig:hfunc}).

Note that the result~(\ref{C4AAres}) is not applicable for $m_{\pi}$ equal or near $m_{K}$. The reason for this is that we omitted the three-kaon contribution to the correlator, because we assumed the kaon to be significantly heavier than the pion.  Because of this Eq.~(\ref{C4AAres}) does not reproduce Eq.~(\ref{C1AAres}) for  $m_K=m_\pi$. 

As a technical aside we mention that diagrams \ref{fig:diagrams}b--d vanish in the calculation of Eq.~(\ref{C1AAres}). The reason is isospin symmetry 
and the fact that all pions have vanishing spatial momenta. 
The particular form of the current~(\ref{A1NLO}) implies that diagrams \ref{fig:diagrams}b--d are
proportional to energy differences of the pions and thus vanish if all pions have equal spatial momenta. Therefore, only diagram~\ref{fig:diagrams}e contributes to $C^1_{AA,3\pi}(t)$.

We emphasize that the three-pion contribution in Eq.~(\ref{C1AAres}) is completely determined in terms of the leading one-pion contribution:  Only  $f_{\pi}$ and $m_{\pi}$ appear as non-trivial, unknown parameters. 
In other words, employing Eq.~(\ref{C1AAres}) in the analysis of actual lattice data leads essentially to a two-parameter fit, involving the same unknowns as a standard single-exponential fit.
The same applies to the kaon correlator once $f_{\pi}$ and $m_{\pi}$ have been determined. 

\subsection{\label{PPresults} Results for the pseudo-scalar correlator}
The pseudo-scalar correlators are easily obtained using the PCAC relations~(\ref{pcac1}). For flavor index $a=1$ we obtain
\begin{equation}
\label{C1PPres}
C^1_{PP}(t) =  - \frac{f^2B^2}{2m_{\pi}}\; e^{-m_{\pi} t}\left[1+ \frac{45}{512 (f_{\pi}L)^4(m_{\pi} L)^2}\;e^{-2m_{\pi}t}\right] \ .
\end{equation}
The three-pion contribution is enhanced by a factor 9 compared to Eq.~(\ref{C1AAres}). This is a simple consequence of the two time derivatives in Eq.~(\ref{relcorr}) acting on the exponential $\exp(-3m_{\pi}t)$. These result in a factor $9m_{\pi}^2$ with the $9$ multiplying the factor $5/512$. 

The kaon correlator is obtained analogously. In this case the enhancement factor is given by the ratio $(m_{K}+2m_{\pi})^2/m_{K}^2$, so we find
\begin{equation}
\label{C4PPres}
C^4_{PP}(t) = 
- \frac{f_K^2B^2}{2m_{K}}e^{-m_{K}t}\Bigg[1+ \frac{1}{24 (f_KL)^4(m_{\pi}L)^2} \,h_2\left(\frac{m_{\pi}}{m_{K}}\right) e^{-2m_{\pi}t}\Bigg]\ .
\end{equation}
The function $h_2(x)$ is related to the previously defined $h_1(x)$ by
\begin{equation}
\label{Defhtilde}
h_2(x)= (1+2x)^2 h_1(x)\ ,
\end{equation}
and sketched in Fig.~\ref{fig:hfunc}.

In principle, the enhancement up to a factor of 9 of the pseudo-scalar correlator gives some preference to the axial-vector correlator in the determination of the mass and decay constant of the pion and kaon. However, the three-particle contribution is very small for both correlators so that it does not play a role in practice in any case.

\subsection{\label{DsMeson} Partially quenched mesons with one or two heavy quarks}
So far we considered 2+1 flavor QCD and ChPT. Useful extensions are the partially quenched (PQ) variants in which the masses of some of the valence
quarks are chosen to be different from any of the sea quark masses \cite{Bernard:1993sv}. Theoretically, these theories have a larger field content due to the presence of additional ghost fields. Although technically more involved, the calculation of the previous section can in principle be repeated for the PQ case. A full PQ calculation is beyond the scope of this paper, but for this subsection we only need to consider two special cases. 

First, we claim that the result~(\ref{C4AAres}) for the kaon correlator applies without change to the $N_f=2$ theory with two light quark flavors and a quenched strange quark. Quenching of the strange quark is achieved by introducing a ghost quark with mass $m_s$ which exactly cancels the contribution of the strange quark to the fermion determinant. In PQChPT the larger field content appears explicitly because the non-linear $\Sigma$ matrix of Eq.~(\ref{Sigma}) becomes a $4\times4$ matrix. The mass matrix in Eq.~(\ref{Mmatrix}) changes to $M={\rm diag}(m,m,m_s,m_s)$. 

Now suppose we are interested in the $\exp[-(m_K + 2m_{\pi})t]$ contribution to the axial-vector correlation function $C^4_{AA}$. As before we can simplify the calculation of the diagrams \ref{fig:diagrams}b--e by dropping terms in the axial-vector current and interaction vertices that do not contribute to the exponential we are interested in. Doing this we end up with the same expressions as in Eqs.~(\ref{A4NLO}) and~(\ref{KaonV}): The three-pion term in the axial-vector current needs to contain at least one $\pi^4$ field and two more light pion fields. There are no other terms than those already shown in Eq.~(\ref{A4NLO}) fulfilling this requirement, with or without the ghost fields. The same is true for the vertices. Three of the four pion fields in the vertices need to be one $\pi_4$ and two light fields, and Eq.~(\ref{KaonV}) contains all of those.
We conclude that result~(\ref{C4AAres}) for the 2+1 flavor theory applies to the theory in which the strange quark is quenched as well.

Next, we want to generalize our results to a partially quenched setup with two light flavors and two quenched heavy flavors. We will refer to  these heavy flavors as strange and charm, even though we consider the case with equal quark masses only, taking $m_c=m_s$. This theory contains a ``$D_s$'' meson, made of the strange and charm quarks, even though with our choice for the quark masses its mass is not the physical $D_s$ meson mass. Nevertheless, we are interested in the axial-vector correlation function that projects on this $D_s$ meson.

The full PQ theory would contain two ghost fields compensating the effects of charm and strange in the fermion determinant. In the corresponding chiral effective theory the field $\Sigma$ becomes a $6\times 6 $ matrix. However, applying the same arguments as before, one can convince oneself that all the additional fields due to the presence of the ghosts do not contribute to the 
terms in the correlator we are interested in. It is therefore sufficient to consider standard ChPT with four flavors in order to calculate these terms. In this theory the flavor index $a$ in Eq.~(\ref{Sigma}) runs from 1 to 15 and the mass matrix (again) assumes the form $M={\rm diag\,}(m,m,m_s,m_s)$. For the group generators $T^a$ we take the obvious generalization of the Gell-Mann matrices to the group $SU(4)$. 

Expanding the chiral lagrangian in powers of the pseudo-scalar fields we obtain, in addition to the masses in Eq.~(\ref{LOMasses}), 
\begin{equation}
m_{a}^2=
\left\{\begin{array}{rclcl}
B(m+m_s) & \equiv & m_K^2\ , & \quad & a=9,\ldots,12\ ,\\
2Bm_s & \equiv&  m_{D_s}^2\ , & \quad & a=13,14\ ,\\
\frac{2}{3}B(m+5m_s) & \equiv& m_{\tilde{\eta}}^2\ ,& \quad & a=15\ .
\end{array}
\right.\label{LOMassesSU4}
\end{equation}
Note that with our choice for the quark masses we have the relation
\begin{equation}
\label{mpimD}
m^2_{D_s} = 2 m^2_{K}-m^2_{\pi}
\end{equation}
between the LO masses.

The axial-vector correlation function that projects on the $D_s$ meson is the one with flavor index $a=13$ or $14$. To LO we find the results already given in Eqs.~(\ref{ACorrLO}) and~(\ref{PCorrLO}) for the correlators. The dominant three-meson contribution is the one with two kaons (defined as a meson made out of either the strange or charm quark and one of the light
quarks) and one pion, as one can easily see from the quark-flow equivalents of Fig.~\ref{fig:diagrams}. The $D_s+ 2 \pi$ state, on the other hand, does not couple to $C^{13}_{AA}$ to the order we are working here. Calculating  diagrams \ref{fig:diagrams}b--e and keeping only the contributions with the exponential fall off $\exp[-(2m_{K}+m_{\pi})t]$ we find:
\begin{equation}
\label{C13AAres}
C^{13}_{AA}(t) = 
\frac{f_{D_s}^2m_{D_s}}{2}e^{-m_{D_s}t}\Bigg[1+ \frac{1}{3 (f_{\pi} L)^4(m_{\pi}L)^2}\frac{m_{\pi}}{m_{D_s}}h_3\left(\frac{m_{\pi}}{m_{K}}\right) e^{-(2m_K +m_{\pi}-m_{D_s})t}\Bigg]\ ,
\end{equation}
where the function $h_3(x)$ is given by
\begin{equation}
\label{Defh3}
h_3(x)= \left(1-x -\frac{2-4x-2x^2+x^3}{4(1+x)^2}\right)^2\ .
\end{equation}
As before, the pseudo-scalar density correlator is easily obtained from Eq.~(\ref{C13AAres}) using the PCAC relation. Here the enhancement factor of the three-meson contribution is $(2m_{K}+m_{\pi})^2/m_{D_s}^2$, so we find
\begin{equation}
\label{C13PPres}
C^{13}_{PP}(t) = 
-\frac{f_D^2}{2m_{D_s}}e^{-m_{D_s}t}\Bigg[1+ \frac{1}{3 (f_{\pi} L)^4(m_{\pi}L)^2}\frac{m_{\pi}}{m_{D_s}}h_4\left(\frac{m_{\pi}}{m_{K}}\right) e^{-(2m_K +m_{\pi}-m_{D_s})t}\Bigg]\ ,
\end{equation}
with $h_4(x)$ defined by
\begin{equation}
\label{Defh4}
h_4(x)=\frac{(2+x)^2}{2-x^2}\;h_3(x)\ .
\end{equation}

\begin{figure}[tp]
\begin{center}
\includegraphics[scale=0.7]{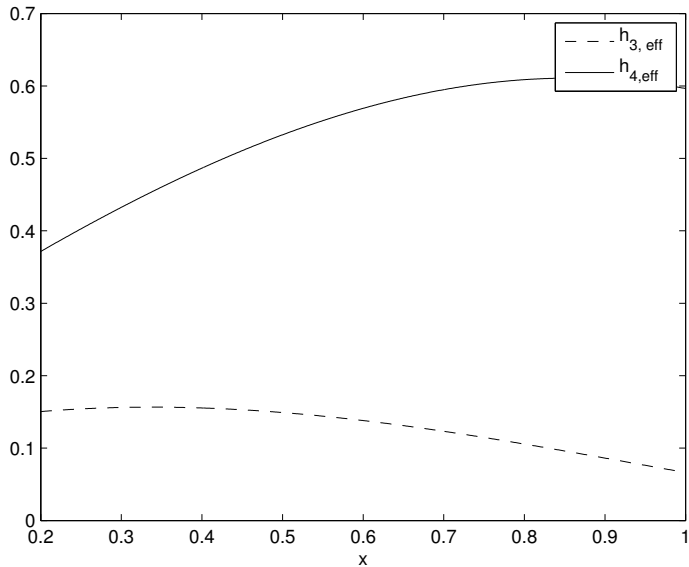}
\caption{The functions $h_{i,{\rm eff}}(x), \,i=3,4$ that enter the results for $C^{13}_{AA}(t)$ and $C^{13}_{PP}(t)$.}
\label{fig:hfunc2}
\end{center}
\end{figure}
Comparing the results~(\ref{C13AAres}) and~(\ref{C13PPres}) with their counterparts~(\ref{C4AAres}) and~(\ref{C4PPres}) for the kaon correlator there appears to be an enhancement factor 8 for the three-meson contribution ($1/3$ versus $1/24$). However, this is not the case when one takes into account all factors.  Incorporating the factor $m_\pi/m_{D_s}=\sqrt{x^2/(2-x^2)}, x=m_{\pi}/m_{K}$, Fig.~\ref{fig:hfunc2} shows the functions 
\begin{equation}
\label{Defhieff}
h_{i,{\rm eff}}(x) = \sqrt{\frac{x^2}{2-x^2}}\,h_i(x),\qquad i=3,4\,,
\end{equation}
that enter the results for the correlators. These functions are roughly a factor 4 smaller than the ones shown in Fig.~\ref{fig:hfunc}, effectively reducing the enhancement to a factor of about 2. 

\subsection{\label{NegContr} A comment on higher-order contributions}
\begin{figure}[tp]
\begin{center}
\includegraphics[scale=0.35]{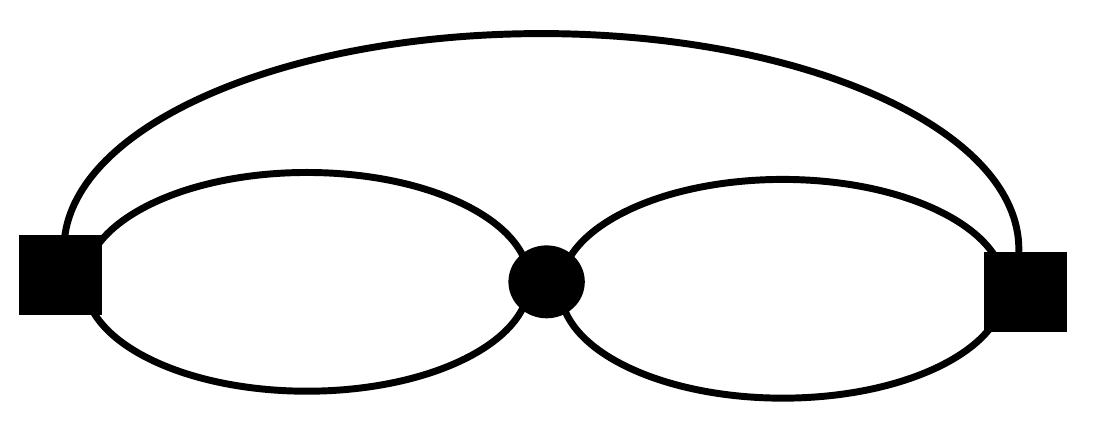}
\caption{Higher-order Feynman diagram for the axial-vector correlation function. }
\label{fig:hodiagram}
\end{center}
\end{figure}

We have assumed that we can take the energy of the three-particle state to be equal to just the sum of the individual masses, as would be the case if the three mesons would not interact. The pseudo-Goldstone boson character of the pions implies that they  interact only weakly if all three pions have small (or vanishing) spatial momenta. Interaction corrections thus appear only at higher order in the chiral expansion and are suppressed by additional inverse powers of the volume. Figure \ref{fig:hodiagram} shows an example of a higher order diagram, generated by a vertex insertion into diagram \ref{fig:diagrams}b. It consists of two more propagators and one more vertex compared to diagram~\ref{fig:diagrams}b. According to the counting rules of Sec.~\ref{2piCont}, it is suppressed by an additional factor $1/L^3$. The time integration leads to term with time dependence $\sim t\exp(-3m_{\pi}t)$, which, after exponentiating, corresponds to the expected energy shift of order $L^{-3}$ 
\cite{Luscher:1986pf}. To the order we are working here such
interaction corrections can be ignored.  

\newpage
\subsection{\label{NegContr2} An estimate for contributions with non-zero momentum pions}
In Section~\ref{secQCD} we assumed that the three-particle contribution to the correlators is dominated by the state with all three mesons at rest, \ceef\ Eqs.~(\ref{tpzerop}) and~(\ref{tpcontr}). For sufficiently large spatial extent $L$ (at fixed pion mass) this is not justified because the ``non-zero momentum states,'' \ie, states where some of the pions have non-zero spatial momentum, are no longer well separated from the one with all momenta equal to zero. 
The calculation of the contributions due to the non-zero momentum states is beyond the scope of the present paper. However, a simple estimate for this contribution will be useful in the next section. 

Let us consider the first subleading contribution in Eq.~(\ref{tpcontr}) with two of the three pions having opposite momenta with magnitude $2\pi/L$. The energy of such a pion can be written as $s_{p}m_{\pi}$ with the scaling factor 
\begin{equation}
\label{sp}
s_{p}=\sqrt{1+\frac{4\pi^2}{(m_{\pi}L)^2}}\ ,
\end{equation}
hence  $E_{\rm tot} = m_{\pi}(2s_p+1)$ and $8E_{\vp} E_{\vq} E_{\vk}=8m^3_{\pi}s_p^2$ for the contribution of this state to Eq.~(\ref{tpcontr}).
Defining $r(t)$ as the ratio of this contribution to the dominant one in Eq.~(\ref{tpzerop}) we obtain the simple estimate
\begin{equation}
\label{Defrt}
r(t)= \frac{18}{s_p^2}\,e^{-2m_{\pi}(s_p-1)t}\ .
\end{equation}
The 18 in the numerator is the multiplicity of the non-zero momentum state, counting the number of different
ways two non-zero and opposite momenta can be distributed among the three pions.
As mentioned before, making $L$ larger at fixed $m_{\pi}$ the ratio $r$ will eventually grow and the non-zero momentum state contribution cannot be ignored.

In case of the kaon correlator ($a=4$) there is a second state where one non-zero momentum is carried by the kaon.  This leads to a somewhat different estimate, but for order of magnitude estimates Eq.~(\ref{Defrt}) will be sufficient.

\section{\label{sec:data} Confronting lattice data}
Two collaborations \cite{DelDebbio:2006cn,DelDebbio:2007pz,Fritzsch:2012wq} provide sufficient information  in order to compare our results with data. 
In these references plots are shown for the effective mass associated with pseudo-scalar correlation function, with the effective mass defined by
\begin{equation}
\label{DefEffMass}
m_{a,{\rm eff}}(t) = -\frac{d}{dt} \log C^a_{PP}(t)\ .
\end{equation}
If only the ground state contributed to the correlator, the effective mass would be equal to the ground state mass $m_a$, and hence be constant in time. Any deviation from a constant value originates from excited and multi-particle states. Our results for the three-particle contribution can be written in the generic form
\begin{equation}
\label{GenMeff}
m_{a,{\rm eff}}(t) =  m_a\left( 1+k_a\,e^{-(m_a'-m_a)t}\right)\ ,
\end{equation}
with $m_a$ and $m_a'$ denoting the ground state and three-meson state masses, respectively, and a coefficient $k_a$, determined by the results of the previous section.

\begin{figure}[t]
\begin{center}
\includegraphics[scale=0.75]{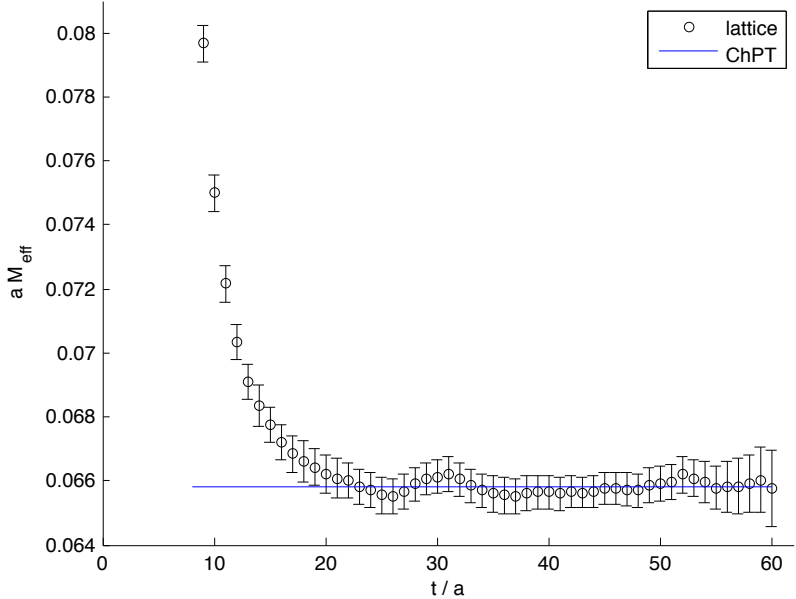}
\caption{The effective mass for the O7 data of Ref.~\cite{Fritzsch:2012wq} (also shown in Fig.~2 of this reference.) The solid blue line corresponds to the ChPT predicition for the three-pion state contribution to the effective mass.}
\label{fig:alphadat}
\end{center}
\end{figure}

Consider first the data reported in Ref.~\cite{Fritzsch:2012wq}. These data were generated with $N_f=2$ O($a$) improved Wilson fermions. One data set, labeled by ``O7,'' has a pion mass of 270~MeV and a spatial extent $L$ such that $m_{\pi}L=4.21$.\footnote{The results we need are found in Tables 1 and 18 of Ref.~\cite{Fritzsch:2012wq}.}  The pion mass is sufficiently light that we may expect ChPT to be applicable.
The measured pseudo-scalar decay constant is such that  $f_{\pi}L=1.58$.\footnote{$Z_A$ necessary to obtain this value can be obtained with Eq.~(B.1) of Ref.~\cite{Fritzsch:2012wq}. Note that we converted $f_{\pi}L$ to our convention in which its physical value is $92.2$~MeV.} The combinations $m_{\pi}L$ and $f_{\pi}L$ are the only unknowns entering the coefficient $k_1$ in the pionic effective mass ($a=1$), 
\begin{equation}
\label{resc1a}
k_1= \frac{90}{512 (f_{\pi}L)^4(m_{\pi}L)^2}\ .
\end{equation}
With the particular values given above we obtain
\begin{equation} 
\label{Alphak1}
k_1\approx 1.6\times 10^{-3}\ .
\end{equation}
This is a tiny number, leading to less than a per mille contribution to the effective mass. Figure~\ref{fig:alphadat} shows the effective mass for the O7 data set together with the ChPT  prediction given by Eqs.~(\ref{GenMeff}) and~(\ref{Alphak1}).\footnote{We thank Francesco Virotta and Rainer Sommer for providing us with the lattice data for this plot.} The latter appears as a straight horizontal line in this plot.  
Apparently, the statistically significant deviation from a constant value in the effective mass data cannot be attributed to the three-pion state contribution.  

Let us turn to the data from the CERN group \cite{DelDebbio:2006cn,DelDebbio:2007pz}. Figure~2 of Ref.~\cite{DelDebbio:2006cn} shows the effective mass of the pseudo-scalar correlator for two different lattices, labelled $D_2$ and $D_4$, respectively. These lattices were also generated with two flavors of non-perturbatively $O(a)$-improved Wilson fermions. 
The measured values for the pseudo-scalar masses and decay constants are collected in Tables 9 and 10 of Ref.~\cite{DelDebbio:2007pz}. 

For the $D_2$ lattice we focus on the data set with equal sea and valence quark masses.\footnote{Second row for the $D_2$ run in tables 9 and 10 of Ref.~\cite{DelDebbio:2007pz}. $Z_A=0.75$ is given in Ref.~\cite{DellaMorte:2005rd}.} Here we find
$f_{\pi}L=1.0$ and $m_{\pi}L=5.9$. Note that for this data set the pion mass is about 620~MeV and the applicability of ChPT is  questionable. Nevertheless, here we find the value 
\begin{equation}
\label{Cernk1}
k_1\approx 4.4\times10^{-3}\,,
\end{equation}
which is of the same order as Eq.~(\ref{Alphak1}). Again, this value is orders of magnitude too small to explain the curvature in the left panel of Fig.~2 of  Ref.~\cite{DelDebbio:2006cn}. 
Strictly speaking, our calculation for $k_1$ does not exactly correspond to the 
lattice data shown in the left panel of Fig.~2, because it shows PQ data with the average valence quark mass roughly ten percent smaller than the sea quark mass. However, this is a small effect that cannot account for the huge discrepency between the data and our ChPT result.

The right panel of Fig.~2 in Ref.~\cite{DelDebbio:2006cn} shows the result for the effective mass with two heavy valence quarks or, with our nomenclature of Sec.~\ref{DsMeson}, for $m_{13,{\rm eff}}(t)$. The data comes from the $D_4$ run with $f_{\pi}L=0.91$ and $m_{\pi}L=4.1$. The $D_s$ mass is not given in Ref.~\cite{DelDebbio:2007pz},
but the value $m_{D_s}=567$ MeV can be read off from Fig.~2.\footnote{The figure shows the result for $\kappa_r=\kappa_s=0.13590$ \cite{Luscherprivcom}.} The kaon mass $m_K= 502$ MeV for this lattice follows from Eq.~(\ref{mpimD}). Using these values in
\begin{equation}
\label{k13ChPT}
k_{13} = \frac{1}{3 (f_{\pi} L)^4(m_{\pi}L)^2}  \frac{(2 m_{K} + m_{\pi} -m_{D_s})m_{\pi}}{m^2_{D_s}}\, h_4\left(\frac{m_{\pi}}{m_K}\right)
\end{equation}
we obtain the estimate
\begin{equation}
\label{Cernk13}
k_{13}\approx 2.1\times 10^{-2}\,.
\end{equation}
As in the previous cases, this value is too small to account for the curvature in the right panel of Fig.\ 2 of Ref.~\cite{DelDebbio:2006cn}. 

Result \pref{Cernk13} for $k_{13}$ is larger by a factor 4.7 compared to $k_1$ in \pref{Cernk1}. Most of this enhancement stems from the smaller pion mass on the $D_4$ lattice, a smaller part is a partial quenching effect. In order to roughly disentangle both effects let us compute $k_1$ for the $D_4$ lattice:
\begin{equation}
\label{Cernk1D4}
k_1\approx 1.5\times10^{-2}\,.
\end{equation}
This is already 3.4 times larger than Eq.~(\ref{Cernk1}). The remaining factor 1.4 compared to Eq.~(\ref{Cernk13}) can be attributed to the 
valence quark masses being heavier than the sea quark masses.

Finally, let us estimate the contribution of the non-zero momentum states
using Eq.~(\ref{Defrt}). We choose a sufficiently small time separation $t$ where the exponential factor does not completely suppress this contribution and where the data differs significantly from the leading ChPT result. 

For the ALPHA data shown in Fig.~\ref{fig:alphadat} we choose $t/a=14$, leading to $r(t)\approx 1.3$. Thus, the non-zero momentum state contribution is not negligible. Still, even if we allow for a few non-zero momentum states  
their cumulative contribution is still too small to describe the data. The same conclusion can be drawn for the data of the CERN group. If we choose $t=0.7$~fm we obtain  $r(t)\approx 1.3$  (left panel) and $r(t)\approx 0.5$ (right panel).

\section{\label{sec:concl} Concluding remarks}
Our results make it extremely unlikely that a multi-pion state is seen in pseudo-scalar density and axial-vector correlation functions. The overlap of these local operators with multi-pion states is tiny, much too small to be seen in current lattice QCD data. This result is qualitatively in accord with findings in excited state spectroscopy \cite{Dudek:2010wm,Bulava:2010yg,Bulava:2011uk}, where the data showed no indication for the presence of multi-meson states, and extended operators need to be employed in order to uncover such states. 

Of course, it is interesting to speculate on how the dependence of the
effective mass plots of Ref.~\cite{DelDebbio:2006cn} on the (sea) pion mass
might be explained.   Possible excited states are the single-particle
resonance corresponding to the $\pi(1300)$, and, for instance, the
$\rho-\pi$ or $\sigma(500)-\pi$ excited states.

The latter two are again multi-particle states, but will be suppressed by
only a factor linear in the spatial volume, instead of quadratic.   
Therefore, the factors corresponding to $k_a$ in Eq.~(\ref{GenMeff}) will
be enhanced compared to the cases we considered in Sec.~\ref{sec:data}.
For the pseudo-scalar channel, a $\rho-\pi$ two-particle state has
angular momentum $L=1$, and therefore the $\rho$ and $\pi$ mesons
should each have at least one unit of momentum $\pm 2\pi/L$.  
Using lattice data for the vector meson provided in Ref.~\cite{DelDebbio:2007pz}
the two-particle energies are at least about $2$~GeV for both cases
considered in Sec.~\ref{sec:data}, respectively.   This is probably too large to 
fit the time dependence of the effective mass plots in Ref.~\cite{DelDebbio:2006cn}.

Assuming that the overlap with a $\sigma(500)-\pi$ two-particle state 
with the local pseudo-scalar density is
very small, we speculate that the mass $m'_a$ in Eq.~(\ref{GenMeff})
may be a resonance for the simulations of Refs.~\cite{DelDebbio:2006cn,DelDebbio:2007pz}.  From the Fig.~2 of Ref.~\cite{DelDebbio:2006cn}
we estimate $m'_a$ and $k_a$, finding for the left panel
$m'_a\approx  1.7$~GeV and $k_a\approx 1.8$, and for the right panel
$m'_a\approx  1.5$~GeV and $k_a\approx 2.2$.\footnote{These values for
$m_a'$ are 
significantly lower than our estimates for the energies of $\rho-\pi$
states.}  We may fit the dependence on the sea pion mass using the
simple form
\begin{equation}
\label{simple}
m'_a=m'_0+b m_\pi^2\ ,\qquad k_a=k_0+a m_\pi^2\ .
\end{equation}
From such a simple-minded fit we find $m'_a\approx 1.3$~GeV at the physical
pion mass, and the estimates $a\approx-2.0$~GeV$^{-2}$ and $b\approx
1.0$~GeV$^{-1}$.  While we cannot tell from this simple exercise 
whether this is the correct interpretation of the time-dependence of the
effective masses in Ref.~\cite{DelDebbio:2006cn}, these numbers appear to
be quite reasonable, and suggest that the first excited state seen in
Ref.~\cite{DelDebbio:2006cn} might correspond to the $\pi(1300)$.
Of course,  a much more detailed investigation, including multiple
operators and several spatial volumes, would be needed to determine
the nature of the time dependence seen in Ref.~\cite{DelDebbio:2006cn}.

Finally,  we anticipate that conclusions similar to those reported here 
should also hold for other correlaters, such as for instance a nucleon
correlator.   Chiral symmetry relates processes with different numbers
of soft pions, and thus we expect that the contribution from a state with two extra pions at rest in the nucleon channel can also be estimated using
the appropriate chiral effective theory.

\vspace{3ex}
\noindent {\bf Acknowledgments}
\vspace{3ex}

We thank Martin L\"uscher, Stefan Schaefer, Francesco Virotta for correspondence and Rainer Sommer and Hank Thacker for discussions. OB is supported in part by the Deutsche Forschungsgemeinschaft (SFB/TR 09). MG is supported in part by the US Department of Energy
and by the Spanish Ministerio de Educaci\'on, Cultura y Deporte, under program SAB2011-0074.
MG also thanks the Galileo Galilei Institute for Theoretical Physics for
hospitality, and the INFN for partial support.

\end{document}